\documentclass[aps, prb, 10pt, preprint, showpacs, preprintnumbers,
superscriptaddress, amsmath, amssymb]{revtex4-1}
\usepackage[pdftex]{graphicx}
\usepackage[ngerman,english]{babel}
\usepackage[ansinew]{inputenc}
\usepackage{natbib}
\usepackage{color}
\usepackage{siunitx}
\usepackage{calc}
\usepackage[colorlinks=true, pdfstartview=FitV, linkcolor=blue,
citecolor=blue, urlcolor=blue]{hyperref}

\definecolor{green2}{rgb}{.0, .58, 0}

\begin{document}

\title{Imaging magnetic vortex configurations in ferromagnetic nanotubes}

\author{M. Wyss} \affiliation{Department of Physics, University of
  Basel, 4056 Basel, Switzerland}
		
\author{A. Mehlin} \affiliation{Department of Physics, University of
  Basel, 4056 Basel, Switzerland}

\author{B. Gross} \affiliation{Department of Physics, University of
  Basel, 4056 Basel, Switzerland}
  
\author{A. Buchter} \affiliation{Department of Physics, University of
  Basel, 4056 Basel, Switzerland}
  
\author{A. Farhan} \affiliation{Laboratory for Micro- and
  Nanotechnology, Paul Scherrer Institute, 5232 Villigen, Switzerland}
\affiliation{Laboratory for Mesoscopic Systems, Department of
  Materials, ETH Z\"urich, 8093 Z\"urich, Switzerland}

\author{M. Buzzi} \affiliation{Swiss Light Source, Paul Scherrer
  Institute, 5232 Villigen, Switzerland}
	
\author{A. Kleibert} \affiliation{Swiss Light Source, Paul Scherrer
  Institute, 5232 Villigen, Switzerland}
	
\author{G. T\"{u}t\"{u}nc\"{u}oglu} \affiliation{Laboratory of
  Semiconductor Materials, Institute of Materials, School of
  Engineering, \'Ecole Polytechnique F\'ed\'erale de Lausanne,
  1015 Lausanne, Switzerland}
  
\author{F. Heimbach} \affiliation{Lehrstuhl f\"{u}r Physik
  funktionaler Schichtsysteme, Physik Department E10, Technische
  Universit\"{a}t M\"{u}nchen, 85747 Garching, Germany}
	
\author{A. Fontcuberta i Morral} \affiliation{Laboratory of
  Semiconductor Materials, Institute of Materials, School of
  Engineering, \'Ecole Polytechnique F\'ed\'erale de Lausanne, 1015
  Lausanne, Switzerland}
	
\author{D. Grundler} \affiliation{Laboratory of Nanoscale Magnetic
  Materials and Magnonics, Institute of Materials, School of
  Engineering, \'Ecole Polytechnique F\'ed\'erale de Lausanne, 1015
  Lausanne, Switzerland}
	
\author{M. Poggio} \affiliation{Department of Physics, University of
  Basel, 4056 Basel, Switzerland} \email{martino.poggio@unibas.ch}

\begin{abstract} 
  We image the remnant magnetization configurations of CoFeB and
  permalloy nanotubes (NTs) using x-ray magnetic circular dichroism
  photo-emission electron microscopy. The images provide direct
  evidence for flux-closure configurations, including a global vortex
  state, in which magnetization points circumferentially around the NT
  axis.  Furthermore, micromagnetic simulations predict and
  measurements confirm that vortex states can be programmed as the
  equilibrium remnant magnetization configurations by reducing the NT
  aspect ratio.
\end{abstract} 

\maketitle

The study of magnetic nanostructures is motivated by their potential
as elements in dense magnetic memories, logical devices
\cite{cowburn_room_2000}, magnetic sensors
\cite{maqableh_low-resistivity_2012}, and as probes in high-resolution
imaging applications \cite{khizroev_direct_2002,
  poggio_force-detected_2010,campanella_nanomagnets_2011}.  For these
reasons, a number of nanometer-scale geometries have been investigated
both theoretically and experimentally, including magnetic dots, rings,
wires, and tubes.  Of particular interest are nanomagnets with stable
flux-closure magnetization configurations and both fast and
reproducible reversal processes.  In the context of magnetic memory,
speed and reliability are determined by the latter, while ultimate
density can be enhanced by the former.  High density is favored by
flux-closure configurations because they produce minimal stray fields,
thereby minimizing interactions between nearby memory elements
\cite{han_nanoring_2008}.

Ferromagnetic nanotubes (NTs) are a particularly promising morphology,
given their lack of a magnetic core.  At equilibrium, the hollow
magnetic geometry is expected to stabilize vortex-like flux-closure
configurations with magnetization pointing along the NT circumference.
In contrast, in magnetic nanowires (NWs), the exchange energy penalty
for the axial singularity in the center of a vortex configuration
tends to favor non-flux-closure states.  In a NT, the lack of this
axial singularity -- or Bloch point structure -- also allows for a
fast magnetization reversal process that begins with vortices
nucleating at its ends and propagating along its length
\cite{usov_domain_2007, landeros_reversal_2007,
  landeros_equilibrium_2009}.  Theoretical studies of ferromagnetic
NTs have predicted an equilibrium flux-closure configuration, the
so-called global vortex state, as well as other equilibrium
configurations including a uniform axial state and a mixed state
\cite{landeros_equilibrium_2009}.  In a global vortex state, the
entire NT's magnetization is circumferentially aligned, while the
mixed state combines vortex-like ends, minimizing magnetostatic
energy, and an axially aligned center, minimizing exchange energy.
Calculations suggest that for short NTs, opposing vortex states, in
which two vortices of opposing chirality are separated by a N\'eel
domain wall, may also be stable \cite{chen_magnetization_2011}.  The
dependence of the NT's equilibrium magnetization configuration on
geometry, as well as details such as the relative chirality of the
end-vortices, have been considered both analytically and numerically
\cite{chen_equilibrium_2007, chen_magnetization_2010,
  chen_magnetization_2011}.  In particular, it has been predicted that
the most technologically relevant magnetization configuration -- the
global vortex state -- can be programed as the stable remnant
configuration by a small NT aspect ratio.

Experimental evidence for vortex configurations in NTs has so far been
limited to magnetic force microscopy images of single NTs reported by
Li et al \cite{li_template-based_2008}.  There, the authors interpret
a nearly vanishing MFM contrast and a small remnant magnetization as
an indication of a global vortex state.  Streubel et al.\ use x-ray
magnetic dichroism photoemission electron microscopy (XMCD-PEEM) to
investigate rolled-up permalloy (Py) membranes, which are 3-$\mu$m in
diameter \cite{streubel_imaging_2014}.  The authors report azimuthal
domain patterns that are commensurable throughout the windings and
attribute the effect to magnetostatic coupling between windings.
Inter-winding exchange coupling, however, is not present.  Here we use
XMCD-PEEM \cite{kimling_photoemission_2011, jamet_quantitative_2015,
  da_col_observation_2014} to image magnetic configurations in
individual magnetic NTs, which are an order of magnitude smaller.
These NTs are prepared as continuous magnetic shells around
nano-templates, allowing for both magnetostatic and exchange coupling.
We find remnant global vortex states and show that aspect ratio can be
used to program the occurrence of different remnant states, including
mixed, opposing vortex, and global vortex states.

\begin{figure}[t]
  \includegraphics{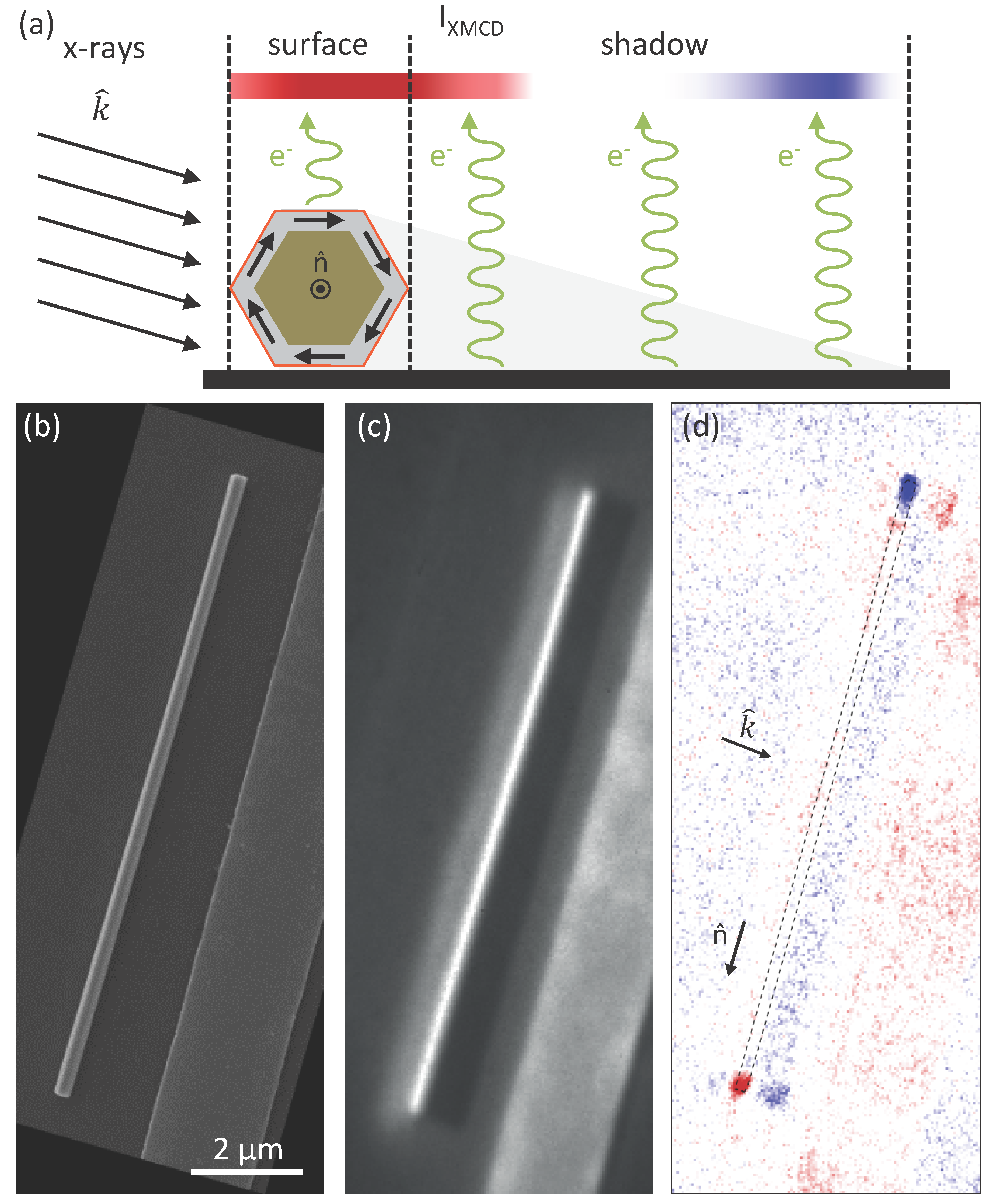}%
  \caption{(a) Schematic drawing of a NT cross-section with incident
    x-rays, photoexcited electrons, and expected XMCD-PEEM contrast
    for the depicted vortex configuration.  The brown central region
    depicts the non-magnetic GaAs template material, the gray region
    the magnetic NT, and the red region the native oxide.  (b) SEM of
    a CoFeB NT with a Au alignment marker visible on the right of the
    image.  (c) PEEM image with grayscale contrast corresponding to
    $I_{\text{PEEM}}$ and (d) XMCD-PEEM image with red (blue) contrast
    representing positive (negative) $I_{\text{XMCD}}$.  The dashed
    line shows the position of the NT.}%
\label{Fig1}
\end{figure}

We study CoFeB and Py NTs consisting of a non-magnetic GaAs core
surrounded by a 30-nm-thick magnetic shell with a hexagonal
cross-section, as shown in Figure~\ref{Fig1}~(a).  The NTs have a
vertex-to-vertex diameters $d$ between 200 and 300 nm and lengths $l$
from 0.5 to 12 $\mu$m.  We obtain specific lengths and well-defined
ends by cutting individual NTs into segments using a focused ion beam
(FIB).  After cutting, we use an optical microscope equipped with
precision micromanipulators to pick up the NT segments and align them
horizontally onto a Si substrate.  Scanning electron micrographs
(SEMs) of the 19 CoFeB and 25 Py NTs studied (see Supporting
Information) reveal continuous surfaces, which are free of detectable
and whose roughness is less than 5 nm.  The fabrication process and
choice of materials avoids magneto-crystalline anisotropy
\cite{hindmarch_interface_2008,ruffer_anisotropic_2014,
  schwarze_magnonic_2013}.  Recent magneto-transport experiments
suggest that a growth-induced magnetic anisotropy may be present in
the CoFeB NTs~\cite{baumgaertl_magnetization_2016}.  Dynamic
cantilever magnetometry measurements of NTs from the same growth
wafers as used here provide $\mu_0 M_S = 1.3 \pm 0.1$ T and $0.8 \pm
0.1$ T for the CoFeB \cite{gross_dynamic_2016} and Py
\cite{buchter_magnetization_2015} NTs, respectively, where $\mu_0$ is
the permeability of free space.  The magnetic shell material is
covered by a few-nanometer-thick native oxide, which affects the NT
magnetization only at cryogenic temperatures through antiferromagnetic
exchange coupling.  In Py NTs, Buchter et al.\ observed such coupling
below a blocking temperature of 18~K
\cite{buchter_magnetization_2015}.

XMCD-PEEM measurements are performed at the Surface/Interface:\
Microscopy (SIM) beamline of the Swiss Light Source (SLS) at the Paul
Scherrer Institut (PSI) \cite{guyader_studying_2012}.  Experiments are
carried out at room temperature and in remnance.  Circularly polarized
x-rays tuned to the $L_{3}$-edge of Fe propagating along $\hat{k}$
impinge on the Si sample substrate with an incident angle of
$16^\circ$, as shown schematically in Figure~\ref{Fig1}~(a).  The
apparatus allows the rotation of the sample about the substrate normal
with respect to $\hat{k}$, which is fixed.  XMCD-PEEM images are
obtained by taking the difference of PEEM images obtained with x-rays
of opposite chirality $\sigma^+$ and $\sigma^-$ normalized to their
sum: $I_{\text{XMCD}} = (I_\sigma^+ - I_\sigma^-)/(I_\sigma^+ +
I_\sigma^-)$, where $I_\sigma^\pm$ represents the emission intensity
of photoelectrons, which is proportional to the local absorption
cross-section of $\sigma^\pm$ polarized x-ray illumination.  The
spatial resolution of the images is about 100 nm.

A SEM of a representative 11.3-$\mu$m-long CoFeB nanotube is shown in
Figure~\ref{Fig1}~(b).  A PEEM image of the same NT shows
$I_{\text{PEEM}} = (I_\sigma^+ + I_\sigma^-)/2$ with the NT long axis
$\hat{n}$ aligned perpendicular to the x-ray beam appears in
Figure~\ref{Fig1}~(c).  Due to the resonant excitation of the Fe
$L_3$-edge, PEEM contrast from the NT surface appears as the brightest
feature.  The dark stripe on the right of the NT is a shadow effect
resulting from the grazing incidence of the incident x-rays and their
partial attenuation by the NT.  A bright region to the left of the NT
appears due to x-rays reflected by the smooth surface of the NT.  A
dotted outline shows the position of the NT in the corresponding
XMCD-PEEM image in Figure~\ref{Fig1}~(d), as determined by overlaying
SEM, PEEM, and XMCD-PEEM images of the same NT.  $I_{\text{XMCD}}$
within the outline stems from a region on top of the NT within 3 to 5
nm of the surface and is proportional to the projection of its local
magnetization along $\hat{k}$.  $I_{\text{XMCD}}$ in the shadow on the
right of the outline -- on the non-magnetic Si surface -- depends on
the imbalance in the intensity of $\sigma^\pm$ x-rays transmitted
through the magnetic material, rather than any surface magnetization
\cite{da_col_observation_2014}.  The contrast in the shadow region is
therefore sensitive to the magnetization of the volume traversed by
the x-rays, with opposite sign compared to surface contrast (see
Methods).  The lack of surface and shadow $I_{\text{XMCD}}$ contrast
corresponding to the central part of the NT in Figure~\ref{Fig1}~(d)
indicate negligible magnetization oriented parallel to $\hat{k}$
(perpendicular to $\hat{n}$).  On the other hand, at the NT ends,
strong surface and shadow contrast indicates magnetization oriented
parallel to $\hat{k}$ (perpendicular to $\hat{n}$).

\begin{figure}[t]
  \includegraphics{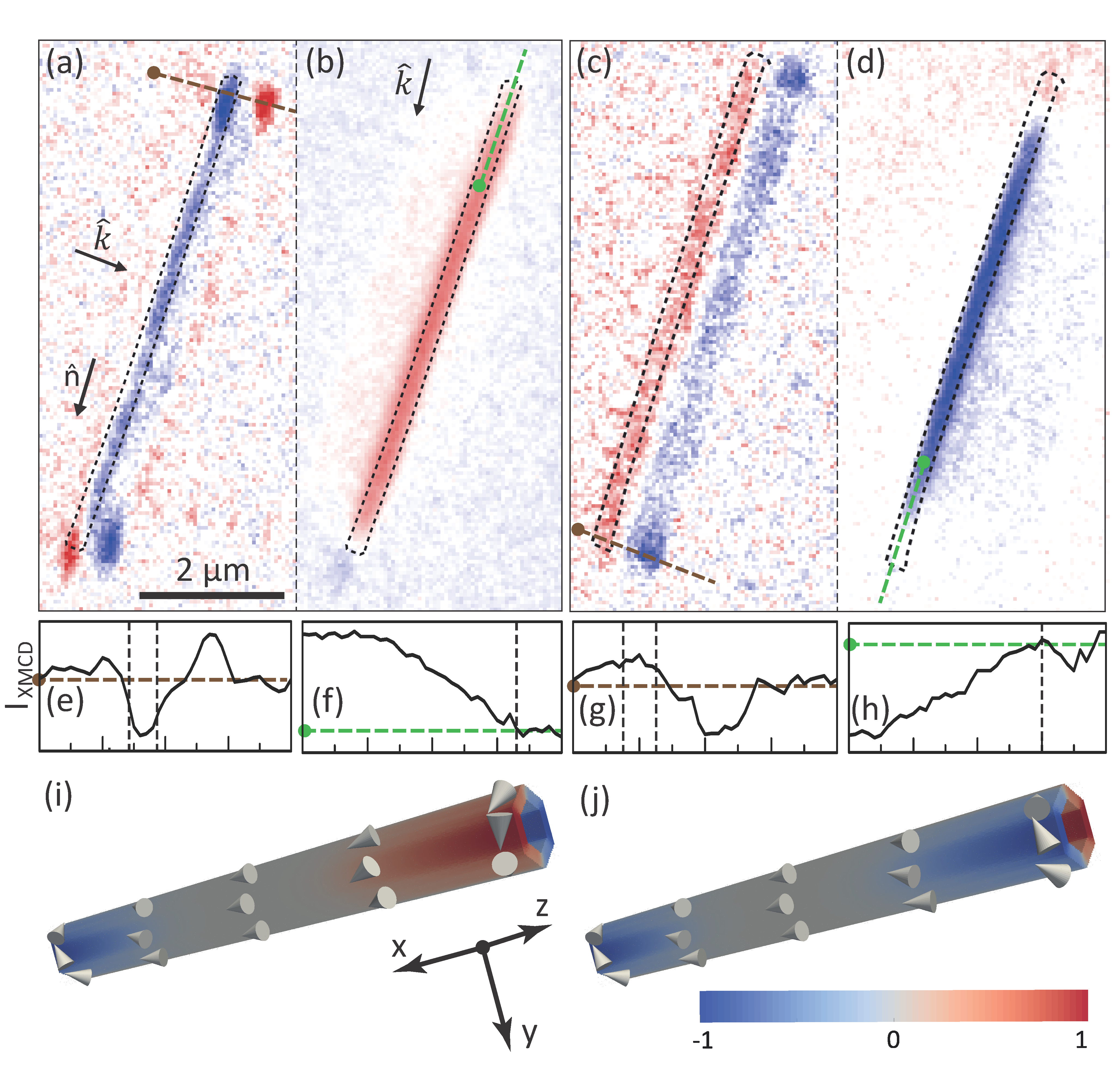}
  \caption{XMCD-PEEM images of a 6.9-$\mu$m-long Py NT with (a)
    $\hat{k} \perp \hat{n}$ and (b) $\hat{k} \parallel \hat{n}$ and of
    a 7.2-$\mu$m-long CoFeB NT with (c) $\hat{k} \perp \hat{n}$ and
    (d) $\hat{k} \parallel \hat{n}$.  Dashed outlines indicate the
    positions of the NTs.  (e) - (h) represent 2-$\mu$m-long
    $I_{\text{XMCD}}$ linecuts along the corresponding colored dashed
    lines in (a) - (d).  In the linecuts, the backround intensity is
    indicated by the level of the horizontal dashed lines and vertical
    dashed lines delinate the boundaries of the NT.  (i) and (j) show
    simulated remnant magnetic states for a NT with $l=2.1$ $\mu$m and
    $d=245$ nm.  Both configurations are mixed states with an axial
    central domain and vortex ends of either (i) opposing chirality --
    consistent with (a) and (b) -- or (j) matching chirality --
    consistent with (c) and (d).  The color-scale corresponds to
    normalized magnetization along $\hat{y}$.  Arrow heads indicate
    the local magnetization direction.}
\label{Fig2}
\end{figure}

XMCD-PEEM images with $\hat{k} \perp \hat{n}$ (perpendicular XMCD-PEEM
contrast) are shown in Figures~\ref{Fig2}~(a) and (c) for Py and CoFeB
NTs with lengths of 6.9 $\mu$m and 7.2 $\mu$m, repectively.  Contrast
similar to that of Figure~\ref{Fig1}~(d) shows magnetization oriented
perpendicular to $\hat{n}$ near the ends of the NTs.  Further
information is gleaned by rotating the sample stage relative to
$\hat{k}$ and performing the same measurements with $\hat{k} \parallel
\hat{n}$ (parallel XMCD-PEEM contrast), as shown in
Figures~\ref{Fig2}~(b) and (d).  In this case, strong surface contrast
in the central part of the NT indicates magnetization parallel to
$\hat{n}$.  Decreased contrast at the NT ends confirms a magnetization
oriented perpendicular to $\hat{n}$, as suggested by the measurements
with $\hat{k} \perp \hat{n}$.  Despite the strong shadow contrast from
the ends in Figure~\ref{Fig2}~(c), the corresponding surface contrast
is weak.  This lack of contrast is likely due to oxidation of the NT
surface.  Given the limited probing depth related to the surface
contrast and the fact that the shadow contrast originates from the
magnetization within the NT, in such cases we rely on
$I_{\text{XMCD}}$ in the shadow to determine the NT's magnetic
configuration.

Specific magnetization configurations in a magnetic nanostructure
produce characteristic XMCD-PEEM signatures for a given orientation of
$\hat{k}$ .  Following the procedure described by Jamet et al., which
takes into account the progressive absorption of the x-ray beam
through the sample cross-section \cite{jamet_quantitative_2015}, a
vortex configuration in our NTs should result in perpendicular
XMCD-PEEM contrast of the form shown in Figure~\ref{Fig1}~(a).  This
contrast is characterized by both strong surface and shadow contrast
(due to the component of the magnetization aligned parallel to
$\hat{k}$) and a change of sign in the x-ray shadow.
Figures~\ref{Fig2}~(e) and (g) show line-cuts through the
perpendicular XMCD-PEEM image at the NT ends that match this
expectation.  Figures~\ref{Fig2}~(f) and (h) show line-cuts of
parallel XMCD-PEEM images through the same region along $\hat{n}$.
The reduction in the surface contrast near the end of the NT relative
to the central region is also consistent with decreasing on-axis
magnetization due to a vortex end state.  The complementarity of
vanishing and strong contrast in the central part of the NT in the
perpendicular and parallel XMCD-PEEM images, respectively, provides
strong evidence for an axially aligned central region.  Taken
together, these images point to a mixed state configuration, where
magnetic moments in the central part of the NT align along its long
axis and curl into vortices at the ends.  The relative signs of the
perpendicular XMCD-PEEM contrast at the ends of the NT shown in
Figure~\ref{Fig2}~(a) (\ref{Fig2}~(c)), indicates that the NT is in a
mixed state with end vortices of opposing (matching) chirality.

\begin{figure}[t]
\includegraphics{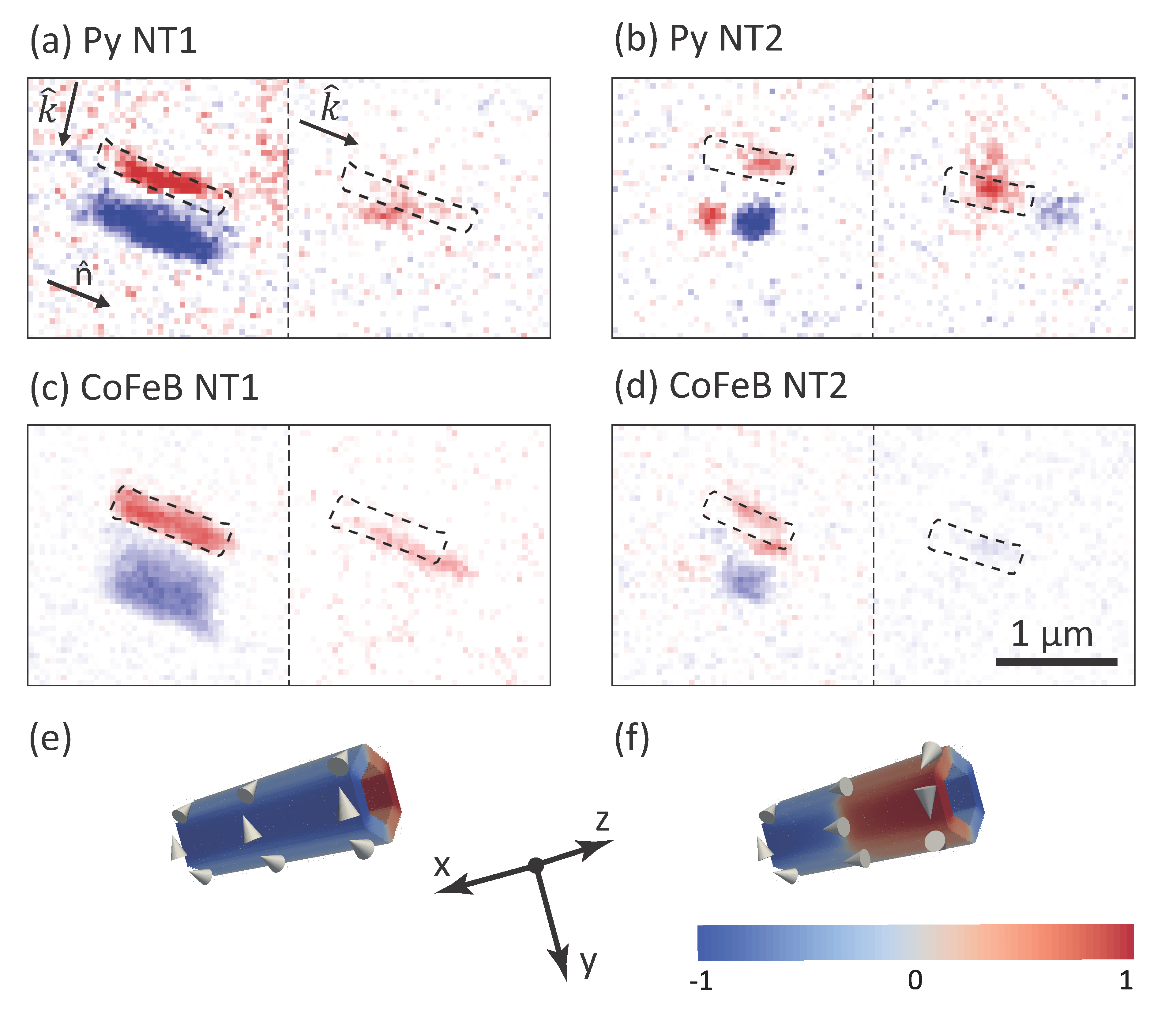}%
\caption{XMCD-PEEM images with $\hat{k} \perp \hat{n}$ and
  $\hat{k} \parallel \hat{n}$ of short NTs.  (a) 1.3-$\mu$m-long Py NT
  found in a global vortex state. (b) 0.73-$\mu$m-long Py NT in an
  opposing vortex state. (c) 1.06-$\mu$m-long CoFeB NT in a global
  vortex state. (d) 0.83-$\mu$m-long CoFeB NT in an opposing vortex
  state. Simulated equilibrium states of ferromagnetic NTs ($l=610$
  nm, $d=245$ nm) in (e) a global vortex state and in (f) an opposing
  vortex state.  The color-scale corresponds to the normalized
  magnetization along $\hat{y}$.  Arrow heads indicate the local
  magnetization direction.}
\label{Fig3}
\end{figure}
 
For NTs of either material longer than 2 $\mu$m, we find remnant mixed
states with vortices of both opposing and matching chirality, as in
Figure~\ref{Fig2} (see Supporting Information).  For NTs shorter than
2 $\mu$m, different magnetization configurations emerge.
Figure~\ref{Fig3}~(a) shows both perpendicular and parallel XMCD-PEEM
images of a 1.30-$\mu$m-long Py nanotube.  In the perpendicular image,
nearly all magnetic moments point perpendicular to $\hat{n}$ and show
the signature of a global vortex state with a single chirality.  In
the parallel image, a small area of axial moments is visible in the
surface contrast, indicating either a slightly tilted vortex
configuration or imperfections at the surface of the magnetic shell.
Figure~\ref{Fig3}~(b) shows XMCD-PEEM contrast from a 0.73-$\mu$m-long
Py NT, in which the magnetization points mostly perpendicular to the
NT axis.  The remnant magnetization configuration, however, does not
display a vortex of a single chirality, but rather two vortices of
opposing chiralities separated by an axial N\'eel domain wall.  This
central wall produces vanishing shadow contrast (white) in the
perpendicular image, while showing strong positive (red) surface and
negative (blue) shadow contrast in the parallel image, as expected for
magnetization aligned along $\hat{n}$.  We therefore conclude that
this NT is in an opposing vortex state.  Results for CoFeB NTs of
similar sizes are shown in Figure~\ref{Fig3}~(c) and (d).
Figure~\ref{Fig3}~(c) shows contrast from a 1.06-$\mu$m-long NT in a
remnant global vortex state, whereas (d) shows a 0.83-$\mu$m-long NT
in a remnant opposing vortex state.

In order to corroborate the XMCD-PEEM measurements, we carry out
numerical simulations of NT magnetization using the software package
\textit{Mumax3} \cite{vansteenkiste_design_2014}.  This package
employs the Landau-Lifshitz micromagnetic formalism using a
finite-difference discretization.  We then compare the numerically
expected equilibrium configurations and the experimentally observed
remnant configurations as a function of vertex-to-vertex diamter $d$
and length $l$.  Experimental results are extracted from perpendicular
and parallel XMCD-PEEM images of each measured NT (see Supporting
Information).  Simulations are consistent with our XMCD-PEEM
measurements in that long NTs are calculated to have remnant mixed
state configurations, as depicted in Figure~\ref{Fig2}~(i) and (j),
and short NTs remnant global vortex or opposing vortex states, as
depicted in Figure~\ref{Fig3}~(e) and (f).  The simulations also
reproduce subtleties of the magnetization configuration in the NTs,
including the length of the vortex ends and its dependence on $d$ (see
Supporting Information).  The average length of the vortices along
$\hat{n}$ is measured to be 320 nm for CoFeB NTs and 360 nm for Py
NTs.

Despite the agreement, the relative chirality of the end vortices
predicted by the simulations does not match our observations.  For
long NTs in remnant mixed state configurations, the energy difference
between a configuration with matching or opposing chirality vortices
is calculated to be small compared to the precision of the simulation;
therefore each is predicted to be equally likely.  In the real NTs,
although the distribution is equal across all NTs, it is unequal for a
single material: 3 opposing and 8 matching mixed states appear in
CoFeB NTs, while 9 opposing and 4 matching appear in Py NTs.  As
aspect ratio is reduced, simulations indicate that the relative
chirality of the end vortices leads to energy differences larger than
the thermal energy.  As the central region of axial magnetization
disappears, opposing chirality should first be favored, resulting in a
stable opposing vortex state.  Upon further reductions in aspect
ratio, the simulations eventually favor matching chirality, resulting
in a stable global vortex state.  For short NTs of both CoFeB and Py,
however, the measured distribution of relative vortex chirality as a
function of $l$ and $d$ does not follow the numerical predictions (see
Supporting Information).  These discrepancies suggest that
imperfections may be decisive in energetically favoring one
configuration over another.  Simulations show that the equilibrium
chirality is sensitive to variations in NT thickness as well as
geometrical imperfections, such as slanted rather than flat ends
introduced by the FIB cutting process.  Given that such imperfections
are known to be present, we assume that they play a role in
determining the relative chirality of end vortices.

\begin{figure}[h!]
\includegraphics{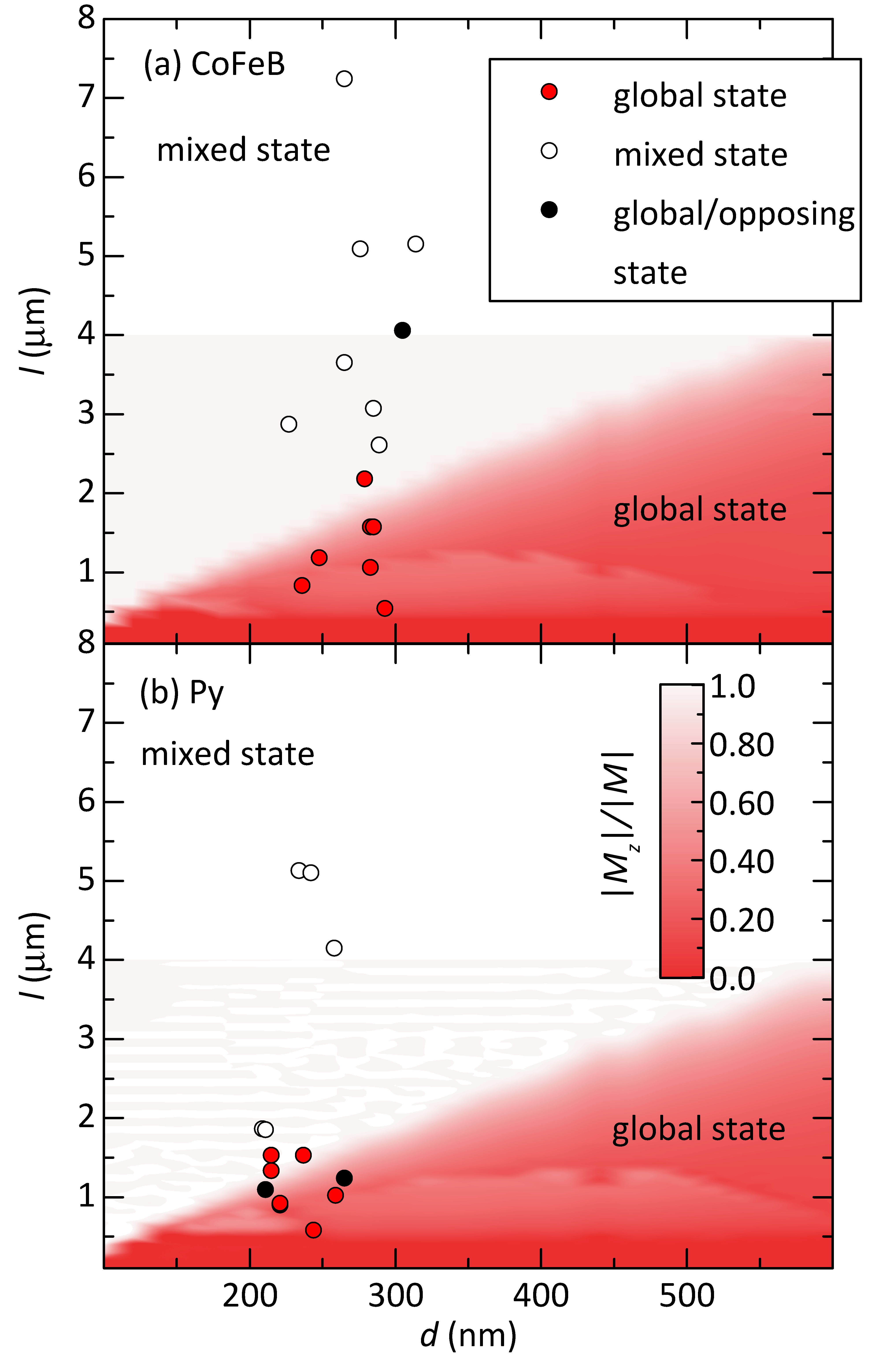}%
\caption{Phase diagrams for (a) CoFeB and (b) Py NTs as a function of
  $l$ and $d$ considering only equal chirality magnetization
  configurations.  The numerically calculated order parameter
  $M_n/|\mathbf{M}|$ is plotted in the color scale and defines the
  boundary between the mixed and global vortex state.  Red (white)
  points show measurements of real NTs in global vortex (mixed) states
  as measured by XMCD-PEEM (see Supporting Information).  Black points
  show NTs that switched remnant configuration after we applied 40 mT
  along $\hat{n}$. }
\label{Fig4}
\end{figure}

If we consider NTs measured to have equal chirality vortices, we find
that the NT aspect ratio dictates whether the remnant state is a
global vortex state, consistent with the simulations.  In
Figure~\ref{Fig4}~(a) and (b), NTs measured to be in either a mixed or
global vortex state are plotted as a function of $l$ and $d$ together
with the numerial expectation.  We distinguish between a global vortex
state and a mixed state with an axially aligned central domain and
equal chirality vortices.  The presence of the central domain can be
quantified by an order parameter $M_n/|\mathbf{M}|$ corresponding to
the relative amount of magnetization that is axially aligned.  By
plotting this order parameter for numerical simulations of NTs with
different lengths and diameters, a clear boundary between the mixed
state and global vortex state emerges, as shown in
Figure~\ref{Fig4}~(a) and (a).  The same classification is carried out
on the measurements, with the global vortex state determined by the
absence of a well-resolved axial domain in the XMCD-PEEM images.  The
measured dependence of magnetic configuration on geometry not only
agrees closely with the numerical simulations, but is also
qualitatively consistent with analytical predictions for cylindrical
ferromagnetic NTs by Landeros et al~\cite{landeros_equilibrium_2009}.

If we consider NTs with opposing chirality, a similar phase boundary
can be defined between a mixed state configuration with opposing end
vortices and an opposing vortex state incorporating a N\'eel wall (see
Supporting Information).  The transition from an axial central domain,
found in the mixed state, to the N\'eel domain wall can be quantified
by the presence of inflection points in plots of $M_n/|\mathbf{M}|$
along $\hat{n}$. Unfortunately, the available spatial resolution of
XMCD-PEEM is not sufficient to clearly determine the inflection points
and therefore to distinguish between these two states.

In order to test the robustness of the remnant magnetization
configurations, for some NTs, we take a second set of XMCD-PEEM images
in remnance after applying 40 mT along $\hat{n}$ \textit{in situ} to
saturate the NTs.  In 11 cases, the measured remnant configuration is
observed to be identical to the one initially measured, while in 6
cases the relative chirality of the end vortices changes.  According
to the simulations, in these 6 cases, the dimensions of the NTs are
such that a matching or opposing chirality does not significantly
affect its magnetic energy (see Supporting Information).  These
include long NTs in mixed states and NTs calculated to be close to the
phase boundary between where opposing or equal chirality is favored
(black points in Figure~\ref{Fig4}~(b)).  Sample imperfections may
trigger the chirality change in such NTs.

In conclusion, we use XMCD-PEEM to image the remnant magnetization
configuration of CoFeB and Py NTs for a variety of lengths.  Our study
reveals that short NTs can occupy a stable global vortex state in
remnance.  Consistent with an analytical theory by Landeros et al.\
\cite{landeros_equilibrium_2009} and our own numerical simulations,
the NT aspect ratio is found to play a crucial role in stabilizing the
global vortex state.  XMCD-PEEM images of the equilibrium
magnetization configurations show that the relative chirality of
vortex ends in real NTs is less controlled than expected from
simulations.  As a result, short NTs are found not only in remnant
global vortex states, but also in opposing vortex states, which
include a N\'eel wall between two opposing vortices.  Additional
simulations suggest that sample imperfections including variations in
thickness and deviations from a perfect geometry are responsible for
this discrepancy.  Still, our magnetic images of global vortex states
show that the most important properties predicted for idealized
ferromagnetic NTs have been realized in real structures.  They also
demonstrate the programming of the equilibrium magnetic configuration
of a ferromagnetic NT via geometry, a result consistent with
long-standing theoretical predictions.

\textbf{Methods.}  \textit{Sample Preparation.}  Ferromagnetic NTs are
made by depositing a thin magnetic film on template GaAs NWs.  These
templates are grown by molecular beam epitaxy on a Si (111) substrate
using Ga droplets as catalysts \cite{ruffer_anisotropic_2014}.  For
the CoFeB NTs, CoFeB is then magnetron-sputtered on the NWs, producing
an amorphous and homogeneous 30-nm thick shell
\cite{gross_dynamic_2016}.  For the Py NTs, a 30-nm thick
polycrystalline shell of Ni$_{80}$Fe$_{20}$ is deposited by thermal
evaporation \cite{buchter_magnetization_2015}.  During both
depositions, the wafers of upright and well-separated GaAs NWs are
mounted with a 35$^\circ$ angle between the long axis of the NWs and
the deposition direction.  The wafers are then continuously rotated in
order to achieve a conformal coating.  

\textit{XMCD-PEEM contrast.}  $I_\sigma^\pm$ at any location is
proportional to both on the intensity of the incident $\sigma^\pm$
x-rays and their absorption at that location.  Absorption of
$\sigma^\pm$ x-rays is proportional to the projection of the magnetic
moment along $\hat{k}$.  Therefore, positive (red) or negative (blue)
$I_{\text{XMCD}}$ represents near surface magnetization either
parallel or anti-parallel with $\hat{k}$, respectively.  For
photoemission excited by x-rays that have previously passed through
magnetic material, however, the absorbtion in the traversed volume
must also be considered \cite{kimling_photoemission_2011,
  jamet_quantitative_2015, da_col_observation_2014}.  In our images,
such magnetic contrast appears in the x-ray shadow of the NT on the
non-magnetic substrate.  Since the absorption of $\sigma^\pm$ x-rays
is proportional to the projection of the magnetic moment along
$\hat{k}$, there is also a proportional attenuation of $\sigma^\pm$
x-rays transmitted through the NT and incident in the shadow.  The
resulting $I_\sigma^\pm$ is therefore proportional to the
magnetization of the volume traversed by the x-rays.  This
proportionality has opposite sign compared to that at a magnetic
surface, i.e.\ positive (red) or negative (blue) $I_{\text{XMCD}}$
results from volume magnetization either anti-parallel or parallel to
$\hat{k}$, respectively.  Combining these two types of contrast, we
extract information about both the surface and volume magnetization of
the measured NTs \cite{kimling_photoemission_2011}.  The roughly 100
nm spatial resolution of the XMCD-PEEM images depends of the quality
of the focus and properties of the sample, including morphology and
cleanliness.

\textit{Mumax3 simulations.}  We set $\mu_0 M_S$ to its measured value
of $\SI{1.3}{\tesla}$ and $\SI{0.8}{\tesla}$ and the exchange
stiffness to $A_{ex} = \SI{28}{\pico\joule}/\si{\meter}$ and
$\SI{13}{\pico\joule}/\si{\meter}$ for CoFeB and Py, respectively.  In
the simulations, space is discretized to $\SI{5}{\nano\meter}$ and
thermal fluctuations are not included.

\textit{Supplementary Information
  (\href{http://poggiolab.unibas.ch/full/WyssSuppInfo.pdf}{WyssSuppInfo.pdf}).}
This file included XMCD-PEEM and SEM images of all measured NTs, phase
diagrams of relative vortex chirality, phase diagrams of NTs with
opposing vortex chirality showing the phase transition between mixed
and opposing vortex states, and plots of both simulated and measured
vortex length as a function of NT diameter.

\acknowledgments The authors thank Jaianth Vijayakumar, David
Bracher, and Patrick Appel for technical support.  We acknowledge the
support of the Canton Aargau, the Swiss Nanoscience Institute, the SNF
under Grant No. 200020-159893, the NCCR Quantum Science and Technology
(QSIT), and DFG in the Schwerpunkt Programm ``Spincaloric transport
phenomena'' SPP1538 via project GR1640/5-2.


\begin{thebibliography}{25}%
\makeatletter
\providecommand \@ifxundefined [1]{%
 \@ifx{#1\undefined}
}%
\providecommand \@ifnum [1]{%
 \ifnum #1\expandafter \@firstoftwo
 \else \expandafter \@secondoftwo
 \fi
}%
\providecommand \@ifx [1]{%
 \ifx #1\expandafter \@firstoftwo
 \else \expandafter \@secondoftwo
 \fi
}%
\providecommand \natexlab [1]{#1}%
\providecommand \enquote  [1]{``#1''}%
\providecommand \bibnamefont  [1]{#1}%
\providecommand \bibfnamefont [1]{#1}%
\providecommand \citenamefont [1]{#1}%
\providecommand \href@noop [0]{\@secondoftwo}%
\providecommand \href [0]{\begingroup \@sanitize@url \@href}%
\providecommand \@href[1]{\@@startlink{#1}\@@href}%
\providecommand \@@href[1]{\endgroup#1\@@endlink}%
\providecommand \@sanitize@url [0]{\catcode `\\12\catcode `\$12\catcode
  `\&12\catcode `\#12\catcode `\^12\catcode `\_12\catcode `\%12\relax}%
\providecommand \@@startlink[1]{}%
\providecommand \@@endlink[0]{}%
\providecommand \url  [0]{\begingroup\@sanitize@url \@url }%
\providecommand \@url [1]{\endgroup\@href {#1}{\urlprefix }}%
\providecommand \urlprefix  [0]{URL }%
\providecommand \Eprint [0]{\href }%
\providecommand \doibase [0]{http://dx.doi.org/}%
\providecommand \selectlanguage [0]{\@gobble}%
\providecommand \bibinfo  [0]{\@secondoftwo}%
\providecommand \bibfield  [0]{\@secondoftwo}%
\providecommand \translation [1]{[#1]}%
\providecommand \BibitemOpen [0]{}%
\providecommand \bibitemStop [0]{}%
\providecommand \bibitemNoStop [0]{.\EOS\space}%
\providecommand \EOS [0]{\spacefactor3000\relax}%
\providecommand \BibitemShut  [1]{\csname bibitem#1\endcsname}%
\let\auto@bib@innerbib\@empty
\bibitem [{\citenamefont {Cowburn}\ and\ \citenamefont
  {Welland}(2000)}]{cowburn_room_2000}%
  \BibitemOpen
  \bibfield  {author} {\bibinfo {author} {\bibfnamefont {R.~P.}\ \bibnamefont
  {Cowburn}}\ and\ \bibinfo {author} {\bibfnamefont {M.~E.}\ \bibnamefont
  {Welland}},\ }\href {\doibase 10.1126/science.287.5457.1466} {\bibfield
  {journal} {\bibinfo  {journal} {Science}\ }\textbf {\bibinfo {volume}
  {287}},\ \bibinfo {pages} {1466} (\bibinfo {year} {2000})}\BibitemShut
  {NoStop}%
\bibitem [{\citenamefont {Maqableh}\ \emph {et~al.}(2012)\citenamefont
  {Maqableh}, \citenamefont {Huang}, \citenamefont {Sung}, \citenamefont
  {Reddy}, \citenamefont {Norby}, \citenamefont {Victora},\ and\ \citenamefont
  {Stadler}}]{maqableh_low-resistivity_2012}%
  \BibitemOpen
  \bibfield  {author} {\bibinfo {author} {\bibfnamefont {M.~M.}\ \bibnamefont
  {Maqableh}}, \bibinfo {author} {\bibfnamefont {X.}~\bibnamefont {Huang}},
  \bibinfo {author} {\bibfnamefont {S.-Y.}\ \bibnamefont {Sung}}, \bibinfo
  {author} {\bibfnamefont {K.~S.~M.}\ \bibnamefont {Reddy}}, \bibinfo {author}
  {\bibfnamefont {G.}~\bibnamefont {Norby}}, \bibinfo {author} {\bibfnamefont
  {R.~H.}\ \bibnamefont {Victora}}, \ and\ \bibinfo {author} {\bibfnamefont
  {B.~J.~H.}\ \bibnamefont {Stadler}},\ }\href {\doibase 10.1021/nl301610z}
  {\bibfield  {journal} {\bibinfo  {journal} {Nano Lett.}\ }\textbf {\bibinfo
  {volume} {12}},\ \bibinfo {pages} {4102} (\bibinfo {year} {2012})}\BibitemShut {NoStop}%
\bibitem [{\citenamefont {Khizroev}\ \emph {et~al.}(2002)\citenamefont
  {Khizroev}, \citenamefont {Kryder}, \citenamefont {Litvinov},\ and\
  \citenamefont {Thompson}}]{khizroev_direct_2002}%
  \BibitemOpen
  \bibfield  {author} {\bibinfo {author} {\bibfnamefont {S.}~\bibnamefont
  {Khizroev}}, \bibinfo {author} {\bibfnamefont {M.~H.}\ \bibnamefont
  {Kryder}}, \bibinfo {author} {\bibfnamefont {D.}~\bibnamefont {Litvinov}}, \
  and\ \bibinfo {author} {\bibfnamefont {D.~A.}\ \bibnamefont {Thompson}},\
  }\href {\doibase 10.1063/1.1508164} {\bibfield  {journal} {\bibinfo
  {journal} {Appl. Phys. Lett.}\ }\textbf {\bibinfo {volume} {81}},\
  \bibinfo {pages} {2256} (\bibinfo {year} {2002})}\BibitemShut {NoStop}%
\bibitem [{\citenamefont {Poggio}\ and\ \citenamefont
  {Degen}(2010)}]{poggio_force-detected_2010}%
  \BibitemOpen
  \bibfield  {author} {\bibinfo {author} {\bibfnamefont {M.}~\bibnamefont
  {Poggio}}\ and\ \bibinfo {author} {\bibfnamefont {C.~L.}\ \bibnamefont
  {Degen}},\ }\href {\doibase 10.1088/0957-4484/21/34/342001} {\bibfield
  {journal} {\bibinfo  {journal} {Nanotechnology}\ }\textbf {\bibinfo {volume}
  {21}},\ \bibinfo {pages} {342001} (\bibinfo {year} {2010})}\BibitemShut {NoStop}%
\bibitem [{\citenamefont {Campanella}\ \emph {et~al.}(2011)\citenamefont
  {Campanella}, \citenamefont {Jaafar}, \citenamefont {Llobet}, \citenamefont
  {Esteve}, \citenamefont {Vázquez}, \citenamefont {Asenjo}, \citenamefont
  {Real},\ and\ \citenamefont {Plaza}}]{campanella_nanomagnets_2011}%
  \BibitemOpen
  \bibfield  {author} {\bibinfo {author} {\bibfnamefont {H.}~\bibnamefont
  {Campanella}}, \bibinfo {author} {\bibfnamefont {M.}~\bibnamefont {Jaafar}},
  \bibinfo {author} {\bibfnamefont {J.}~\bibnamefont {Llobet}}, \bibinfo
  {author} {\bibfnamefont {J.}~\bibnamefont {Esteve}}, \bibinfo {author}
  {\bibfnamefont {M.}~\bibnamefont {V\'azquez}}, \bibinfo {author}
  {\bibfnamefont {A.}~\bibnamefont {Asenjo}}, \bibinfo {author} {\bibfnamefont
  {R.~P.~d.}\ \bibnamefont {Real}}, \ and\ \bibinfo {author} {\bibfnamefont
  {J.~A.}\ \bibnamefont {Plaza}},\ }\href {\doibase
  10.1088/0957-4484/22/50/505301} {\bibfield  {journal} {\bibinfo  {journal}
  {Nanotechnology}\ }\textbf {\bibinfo {volume} {22}},\ \bibinfo {pages}
  {505301} (\bibinfo {year} {2011})}\BibitemShut
  {NoStop}%
\bibitem [{\citenamefont {Han}\ \emph {et~al.}(2008)\citenamefont {Han},
  \citenamefont {Wen},\ and\ \citenamefont {Wei}}]{han_nanoring_2008}%
  \BibitemOpen
  \bibfield  {author} {\bibinfo {author} {\bibfnamefont {X.~F.}\ \bibnamefont
  {Han}}, \bibinfo {author} {\bibfnamefont {Z.~C.}\ \bibnamefont {Wen}}, \ and\
  \bibinfo {author} {\bibfnamefont {H.~X.}\ \bibnamefont {Wei}},\ }\href
  {\doibase 10.1063/1.2839774} {\bibfield  {journal} {\bibinfo  {journal}
  {J. Appl. Phys.}\ }\textbf {\bibinfo {volume} {103}},\ \bibinfo
  {pages} {07E933} (\bibinfo {year} {2008})}\BibitemShut {NoStop}%
\bibitem [{\citenamefont {Usov}\ \emph {et~al.}(2007)\citenamefont {Usov},
  \citenamefont {Zhukov},\ and\ \citenamefont {Gonzalez}}]{usov_domain_2007}%
  \BibitemOpen
  \bibfield  {author} {\bibinfo {author} {\bibfnamefont {N.~A.}\ \bibnamefont
  {Usov}}, \bibinfo {author} {\bibfnamefont {A.}~\bibnamefont {Zhukov}}, \ and\
  \bibinfo {author} {\bibfnamefont {J.}~\bibnamefont {Gonzalez}},\ }\href
  {\doibase 10.1016/j.jmmm.2007.02.138} {\bibfield  {journal} {\bibinfo
  {journal} {J. Magn. Magn. Mater.}\ }\textbf
  {\bibinfo {volume} {316}},\ \bibinfo {pages} {255} (\bibinfo {year}
  {2007})}\BibitemShut {NoStop}%
\bibitem [{\citenamefont {Landeros}\ \emph {et~al.}(2007)\citenamefont
  {Landeros}, \citenamefont {Allende}, \citenamefont {Escrig}, \citenamefont
  {Salcedo}, \citenamefont {Altbir},\ and\ \citenamefont
  {Vogel}}]{landeros_reversal_2007}%
  \BibitemOpen
  \bibfield  {author} {\bibinfo {author} {\bibfnamefont {P.}~\bibnamefont
  {Landeros}}, \bibinfo {author} {\bibfnamefont {S.}~\bibnamefont {Allende}},
  \bibinfo {author} {\bibfnamefont {J.}~\bibnamefont {Escrig}}, \bibinfo
  {author} {\bibfnamefont {E.}~\bibnamefont {Salcedo}}, \bibinfo {author}
  {\bibfnamefont {D.}~\bibnamefont {Altbir}}, \ and\ \bibinfo {author}
  {\bibfnamefont {E.~E.}\ \bibnamefont {Vogel}},\ }\href {\doibase
  10.1063/1.2437655} {\bibfield  {journal} {\bibinfo  {journal} {Appl.
  Phys. Lett.}\ }\textbf {\bibinfo {volume} {90}},\ \bibinfo {pages}
  {102501} (\bibinfo {year} {2007})}\BibitemShut {NoStop}%
\bibitem [{\citenamefont {Landeros}\ \emph {et~al.}(2009)\citenamefont
  {Landeros}, \citenamefont {Suarez}, \citenamefont {Cuchillo},\ and\
  \citenamefont {Vargas}}]{landeros_equilibrium_2009}%
  \BibitemOpen
  \bibfield  {author} {\bibinfo {author} {\bibfnamefont {P.}~\bibnamefont
  {Landeros}}, \bibinfo {author} {\bibfnamefont {O.~J.}\ \bibnamefont
  {Suarez}}, \bibinfo {author} {\bibfnamefont {A.}~\bibnamefont {Cuchillo}}, \
  and\ \bibinfo {author} {\bibfnamefont {P.}~\bibnamefont {Vargas}},\ }\href
  {\doibase 10.1103/PhysRevB.79.024404} {\bibfield  {journal} {\bibinfo
  {journal} {Phys. Rev. B}\ }\textbf {\bibinfo {volume} {79}},\ \bibinfo
  {pages} {024404} (\bibinfo {year} {2009})}\BibitemShut {NoStop}%
\bibitem [{\citenamefont {Chen}\ \emph {et~al.}(2011)\citenamefont {Chen},
  \citenamefont {Gonzalez},\ and\ \citenamefont
  {Guslienko}}]{chen_magnetization_2011}%
  \BibitemOpen
  \bibfield  {author} {\bibinfo {author} {\bibfnamefont {A.-P.}\ \bibnamefont
  {Chen}}, \bibinfo {author} {\bibfnamefont {J.~M.}\ \bibnamefont {Gonzalez}},
  \ and\ \bibinfo {author} {\bibfnamefont {K.~Y.}\ \bibnamefont {Guslienko}},\
  }\href {\doibase 10.1063/1.3562190} {\bibfield  {journal} {\bibinfo
  {journal} {J. Appl. Phys.}\ }\textbf {\bibinfo {volume} {109}},\
  \bibinfo {pages} {073923} (\bibinfo {year} {2011})}\BibitemShut {NoStop}%
\bibitem [{\citenamefont {Chen}\ \emph {et~al.}(2007)\citenamefont {Chen},
  \citenamefont {Usov}, \citenamefont {Blanco},\ and\ \citenamefont
  {Gonzalez}}]{chen_equilibrium_2007}%
  \BibitemOpen
  \bibfield  {author} {\bibinfo {author} {\bibfnamefont {A.~P.}\ \bibnamefont
  {Chen}}, \bibinfo {author} {\bibfnamefont {N.~A.}\ \bibnamefont {Usov}},
  \bibinfo {author} {\bibfnamefont {J.~M.}\ \bibnamefont {Blanco}}, \ and\
  \bibinfo {author} {\bibfnamefont {J.}~\bibnamefont {Gonzalez}},\ }\href
  {\doibase 10.1016/j.jmmm.2007.02.132} {\bibfield  {journal} {\bibinfo
  {journal} {J. Magn. Magn. Mater}\ }\textbf
  {\bibinfo {volume} {316}},\ \bibinfo {pages} {e317} (\bibinfo {year}
  {2007})}\BibitemShut {NoStop}%
\bibitem [{\citenamefont {Chen}\ \emph {et~al.}(2010)\citenamefont {Chen},
  \citenamefont {Guslienko},\ and\ \citenamefont
  {Gonzalez}}]{chen_magnetization_2010}%
  \BibitemOpen
  \bibfield  {author} {\bibinfo {author} {\bibfnamefont {A.~P.}\ \bibnamefont
  {Chen}}, \bibinfo {author} {\bibfnamefont {K.~Y.}\ \bibnamefont {Guslienko}},
  \ and\ \bibinfo {author} {\bibfnamefont {J.}~\bibnamefont {Gonzalez}},\
  }\href {\doibase 10.1063/1.3488630} {\bibfield  {journal} {\bibinfo
  {journal} {J. Appl. Phys.}\ }\textbf {\bibinfo {volume} {108}},\
  \bibinfo {pages} {083920} (\bibinfo {year} {2010})}\BibitemShut {NoStop}%
\bibitem [{\citenamefont {Li}\ \emph {et~al.}(2008)\citenamefont {Li},
  \citenamefont {Thompson}, \citenamefont {Bergmann},\ and\ \citenamefont
  {Lu}}]{li_template-based_2008}%
  \BibitemOpen
  \bibfield  {author} {\bibinfo {author} {\bibfnamefont {D.}~\bibnamefont
  {Li}}, \bibinfo {author} {\bibfnamefont {R.~S.}\ \bibnamefont {Thompson}},
  \bibinfo {author} {\bibfnamefont {G.}~\bibnamefont {Bergmann}}, \ and\
  \bibinfo {author} {\bibfnamefont {J.~G.}\ \bibnamefont {Lu}},\ }\href
  {\doibase 10.1002/adma.200801455} {\bibfield  {journal} {\bibinfo  {journal}
  {Adv. Mater.}\ }\textbf {\bibinfo {volume} {20}},\ \bibinfo {pages} {4575}
  (\bibinfo {year} {2008})}\BibitemShut {NoStop}%
\bibitem [{\citenamefont {Streubel}\ \emph {et~al.}(2014)\citenamefont
  {Streubel}, \citenamefont {Han}, \citenamefont {Kronast}, \citenamefont
  {Ünal}, \citenamefont {Schmidt},\ and\ \citenamefont
  {Makarov}}]{streubel_imaging_2014}%
  \BibitemOpen
  \bibfield  {author} {\bibinfo {author} {\bibfnamefont {R.}~\bibnamefont
  {Streubel}}, \bibinfo {author} {\bibfnamefont {L.}~\bibnamefont {Han}},
  \bibinfo {author} {\bibfnamefont {F.}~\bibnamefont {Kronast}}, \bibinfo
  {author} {\bibfnamefont {A.~A.}\ \bibnamefont {\"Unal}}, \bibinfo {author}
  {\bibfnamefont {O.~G.}\ \bibnamefont {Schmidt}}, \ and\ \bibinfo {author}
  {\bibfnamefont {D.}~\bibnamefont {Makarov}},\ }\href {\doibase
  10.1021/nl501333h} {\bibfield  {journal} {\bibinfo  {journal} {Nano Lett.}\
  }\textbf {\bibinfo {volume} {14}},\ \bibinfo {pages} {3981} (\bibinfo {year}
  {2014})}\BibitemShut {NoStop}%
\bibitem [{\citenamefont {Kimling}\ \emph {et~al.}(2011)\citenamefont
  {Kimling}, \citenamefont {Kronast}, \citenamefont {Martens}, \citenamefont
  {Böhnert}, \citenamefont {Martens}, \citenamefont {Herrero-Albillos},
  \citenamefont {Tati-Bismaths}, \citenamefont {Merkt}, \citenamefont
  {Nielsch},\ and\ \citenamefont {Meier}}]{kimling_photoemission_2011}%
  \BibitemOpen
  \bibfield  {author} {\bibinfo {author} {\bibfnamefont {J.}~\bibnamefont
  {Kimling}}, \bibinfo {author} {\bibfnamefont {F.}~\bibnamefont {Kronast}},
  \bibinfo {author} {\bibfnamefont {S.}~\bibnamefont {Martens}}, \bibinfo
  {author} {\bibfnamefont {T.}~\bibnamefont {B\"ohnert}}, \bibinfo {author}
  {\bibfnamefont {M.}~\bibnamefont {Martens}}, \bibinfo {author} {\bibfnamefont
  {J.}~\bibnamefont {Herrero-Albillos}}, \bibinfo {author} {\bibfnamefont
  {L.}~\bibnamefont {Tati-Bismaths}}, \bibinfo {author} {\bibfnamefont
  {U.}~\bibnamefont {Merkt}}, \bibinfo {author} {\bibfnamefont
  {K.}~\bibnamefont {Nielsch}}, \ and\ \bibinfo {author} {\bibfnamefont
  {G.}~\bibnamefont {Meier}},\ }\href {\doibase 10.1103/PhysRevB.84.174406}
  {\bibfield  {journal} {\bibinfo  {journal} {Phys. Rev. B}\ }\textbf {\bibinfo
  {volume} {84}},\ \bibinfo {pages} {174406} (\bibinfo {year}
  {2011})}\BibitemShut {NoStop}%
\bibitem [{\citenamefont {Jamet}\ \emph {et~al.}(2015)\citenamefont {Jamet},
  \citenamefont {Da~Col}, \citenamefont {Rougemaille}, \citenamefont
  {Wartelle}, \citenamefont {Locatelli}, \citenamefont {Mentes},
  \citenamefont {Santos~Burgos}, \citenamefont {Afid}, \citenamefont {Cagnon},
  \citenamefont {Bochmann}, \citenamefont {Bachmann}, \citenamefont
  {Fruchart},\ and\ \citenamefont {Toussaint}}]{jamet_quantitative_2015}%
  \BibitemOpen
  \bibfield  {author} {\bibinfo {author} {\bibfnamefont {S.}~\bibnamefont
  {Jamet}}, \bibinfo {author} {\bibfnamefont {S.}~\bibnamefont {Da~Col}},
  \bibinfo {author} {\bibfnamefont {N.}~\bibnamefont {Rougemaille}}, \bibinfo
  {author} {\bibfnamefont {A.}~\bibnamefont {Wartelle}}, \bibinfo {author}
  {\bibfnamefont {A.}~\bibnamefont {Locatelli}}, \bibinfo {author}
  {\bibfnamefont {T.~O.}\ \bibnamefont {Mente\c{s}}}, \bibinfo
  {author} {\bibfnamefont {B.}~\bibnamefont {Santos~Burgos}}, \bibinfo {author}
  {\bibfnamefont {R.}~\bibnamefont {Afid}}, \bibinfo {author} {\bibfnamefont
  {L.}~\bibnamefont {Cagnon}}, \bibinfo {author} {\bibfnamefont
  {S.}~\bibnamefont {Bochmann}}, \bibinfo {author} {\bibfnamefont
  {J.}~\bibnamefont {Bachmann}}, \bibinfo {author} {\bibfnamefont
  {O.}~\bibnamefont {Fruchart}}, \ and\ \bibinfo {author} {\bibfnamefont
  {J.~C.}\ \bibnamefont {Toussaint}},\ }\href {\doibase
  10.1103/PhysRevB.92.144428} {\bibfield  {journal} {\bibinfo  {journal} {Phys.
  Rev. B}\ }\textbf {\bibinfo {volume} {92}},\ \bibinfo {pages} {144428}
  (\bibinfo {year} {2015})}\BibitemShut {NoStop}%
\bibitem [{\citenamefont {Da~Col}\ \emph {et~al.}(2014)\citenamefont {Da~Col},
  \citenamefont {Jamet}, \citenamefont {Rougemaille}, \citenamefont
  {Locatelli}, \citenamefont {Mentes}, \citenamefont {Burgos}, \citenamefont
  {Afid}, \citenamefont {Darques}, \citenamefont {Cagnon}, \citenamefont
  {Toussaint},\ and\ \citenamefont {Fruchart}}]{da_col_observation_2014}%
  \BibitemOpen
  \bibfield  {author} {\bibinfo {author} {\bibfnamefont {S.}~\bibnamefont
  {Da~Col}}, \bibinfo {author} {\bibfnamefont {S.}~\bibnamefont {Jamet}},
  \bibinfo {author} {\bibfnamefont {N.}~\bibnamefont {Rougemaille}}, \bibinfo
  {author} {\bibfnamefont {A.}~\bibnamefont {Locatelli}}, \bibinfo {author}
  {\bibfnamefont {T.~O.}\ \bibnamefont {Mentes}}, \bibinfo {author}
  {\bibfnamefont {B.~S.}\ \bibnamefont {Burgos}}, \bibinfo {author}
  {\bibfnamefont {R.}~\bibnamefont {Afid}}, \bibinfo {author} {\bibfnamefont
  {M.}~\bibnamefont {Darques}}, \bibinfo {author} {\bibfnamefont
  {L.}~\bibnamefont {Cagnon}}, \bibinfo {author} {\bibfnamefont {J.~C.}\
  \bibnamefont {Toussaint}}, \ and\ \bibinfo {author} {\bibfnamefont
  {O.}~\bibnamefont {Fruchart}},\ }\href {\doibase 10.1103/PhysRevB.89.180405}
  {\bibfield  {journal} {\bibinfo  {journal} {Phys. Rev. B}\ }\textbf {\bibinfo
  {volume} {89}},\ \bibinfo {pages} {180405} (\bibinfo {year}
  {2014})}\BibitemShut {NoStop}%
\bibitem [{\citenamefont {Hindmarch}\ \emph {et~al.}(2008)\citenamefont
  {Hindmarch}, \citenamefont {Kinane}, \citenamefont {MacKenzie}, \citenamefont
  {Chapman}, \citenamefont {Henini}, \citenamefont {Taylor}, \citenamefont
  {Arena}, \citenamefont {Dvorak}, \citenamefont {Hickey},\ and\ \citenamefont
  {Marrows}}]{hindmarch_interface_2008}%
  \BibitemOpen
  \bibfield  {author} {\bibinfo {author} {\bibfnamefont {A.~T.}\ \bibnamefont
  {Hindmarch}}, \bibinfo {author} {\bibfnamefont {C.~J.}\ \bibnamefont
  {Kinane}}, \bibinfo {author} {\bibfnamefont {M.}~\bibnamefont {MacKenzie}},
  \bibinfo {author} {\bibfnamefont {J.~N.}\ \bibnamefont {Chapman}}, \bibinfo
  {author} {\bibfnamefont {M.}~\bibnamefont {Henini}}, \bibinfo {author}
  {\bibfnamefont {D.}~\bibnamefont {Taylor}}, \bibinfo {author} {\bibfnamefont
  {D.~A.}\ \bibnamefont {Arena}}, \bibinfo {author} {\bibfnamefont
  {J.}~\bibnamefont {Dvorak}}, \bibinfo {author} {\bibfnamefont {B.~J.}\
  \bibnamefont {Hickey}}, \ and\ \bibinfo {author} {\bibfnamefont {C.~H.}\
  \bibnamefont {Marrows}},\ }\href {\doibase 10.1103/PhysRevLett.100.117201}
  {\bibfield  {journal} {\bibinfo  {journal} {Phys. Rev. Lett.}\ }\textbf
  {\bibinfo {volume} {100}},\ \bibinfo {pages} {117201} (\bibinfo {year}
  {2008})}\BibitemShut {NoStop}%
\bibitem [{\citenamefont {Rüffer}\ \emph {et~al.}(2014)\citenamefont
  {Rüffer}, \citenamefont {Slot}, \citenamefont {Huber}, \citenamefont
  {Schwarze}, \citenamefont {Heimbach}, \citenamefont {Tütüncüoglu},
  \citenamefont {Matteini}, \citenamefont {Russo-Averchi}, \citenamefont
  {Kovács}, \citenamefont {Dunin-Borkowski}, \citenamefont {Zamani},
  \citenamefont {Morante}, \citenamefont {Arbiol}, \citenamefont {Morral},\
  and\ \citenamefont {Grundler}}]{ruffer_anisotropic_2014}%
  \BibitemOpen
  \bibfield  {author} {\bibinfo {author} {\bibfnamefont {D.}~\bibnamefont
  {R\"uffer}}, \bibinfo {author} {\bibfnamefont {M.}~\bibnamefont {Slot}},
  \bibinfo {author} {\bibfnamefont {R.}~\bibnamefont {Huber}}, \bibinfo
  {author} {\bibfnamefont {T.}~\bibnamefont {Schwarze}}, \bibinfo {author}
  {\bibfnamefont {F.}~\bibnamefont {Heimbach}}, \bibinfo {author}
  {\bibfnamefont {G.}~\bibnamefont {T\"ut\"unc\"uoglu}}, \bibinfo {author}
  {\bibfnamefont {F.}~\bibnamefont {Matteini}}, \bibinfo {author}
  {\bibfnamefont {E.}~\bibnamefont {Russo-Averchi}}, \bibinfo {author}
  {\bibfnamefont {A.}~\bibnamefont {Kov\'acs}}, \bibinfo {author} {\bibfnamefont
  {R.}~\bibnamefont {Dunin-Borkowski}}, \bibinfo {author} {\bibfnamefont
  {R.~R.}\ \bibnamefont {Zamani}}, \bibinfo {author} {\bibfnamefont {J.~R.}\
  \bibnamefont {Morante}}, \bibinfo {author} {\bibfnamefont {J.}~\bibnamefont
  {Arbiol}}, \bibinfo {author} {\bibfnamefont {A.~F.~i.}\ \bibnamefont
  {Morral}}, \ and\ \bibinfo {author} {\bibfnamefont {D.}~\bibnamefont
  {Grundler}},\ }\href {\doibase 10.1063/1.4891276} {\bibfield  {journal}
  {\bibinfo  {journal} {APL Mater.}\ }\textbf {\bibinfo {volume} {2}},\
  \bibinfo {pages} {076112} (\bibinfo {year} {2014})}\BibitemShut {NoStop}%
\bibitem [{\citenamefont {Schwarze}\ and\ \citenamefont
  {Grundler}(2013)}]{schwarze_magnonic_2013}%
  \BibitemOpen
  \bibfield  {author} {\bibinfo {author} {\bibfnamefont {T.}~\bibnamefont
  {Schwarze}}\ and\ \bibinfo {author} {\bibfnamefont {D.}~\bibnamefont
  {Grundler}},\ }\href {\doibase 10.1063/1.4809757} {\bibfield  {journal}
  {\bibinfo  {journal} {Appl. Phys. Lett.}\ }\textbf {\bibinfo {volume}
  {102}},\ \bibinfo {pages} {222412} (\bibinfo {year} {2013})}\BibitemShut
  {NoStop}%
\bibitem [{\citenamefont {Baumgaertl}\ \emph
    {et~al.}(2016)\citenamefont {Baumgaertl}, \citenamefont
    {Heimbach}, \citenamefont {Maendl}, \citenamefont {Rüffer},
    \citenamefont {Fontcuberta~i Morral},\ and\ \citenamefont
    {Grundler}}]{baumgaertl_magnetization_2016}%
  \BibitemOpen \bibfield {author} {\bibinfo {author} {\bibfnamefont
      {K.}~\bibnamefont {Baumgaertl}}, \bibinfo {author}
    {\bibfnamefont {F.}~\bibnamefont {Heimbach}}, \bibinfo {author}
    {\bibfnamefont {S.}~\bibnamefont {Maendl}}, \bibinfo {author}
    {\bibfnamefont {D.}~\bibnamefont {R\"uffer}}, \bibinfo {author}
    {\bibfnamefont {A.}~\bibnamefont {Fontcuberta~i Morral}}, \ and\
    \bibinfo {author} {\bibfnamefont {D.}~\bibnamefont {Grundler}},\
  }\href {\doibase 10.1063/1.4945331} {\bibfield {journal} {\bibinfo
      {journal} {Appl. Phys. Lett.}\ }\textbf {\bibinfo {volume}
      {108}},\ \bibinfo {pages} {132408} (\bibinfo {year}
    {2016})}\BibitemShut {NoStop}%
\bibitem [{\citenamefont {Gross}\ \emph {et~al.}(2016)\citenamefont {Gross},
  \citenamefont {Weber}, \citenamefont {R\"uffer}, \citenamefont {Buchter},
  \citenamefont {Heimbach}, \citenamefont {Fontcuberta~i Morral}, \citenamefont
  {Grundler},\ and\ \citenamefont {Poggio}}]{gross_dynamic_2016}%
  \BibitemOpen
  \bibfield  {author} {\bibinfo {author} {\bibfnamefont {B.}~\bibnamefont
  {Gross}}, \bibinfo {author} {\bibfnamefont {D.~P.}\ \bibnamefont {Weber}},
  \bibinfo {author} {\bibfnamefont {D.}~\bibnamefont {R\"uffer}}, \bibinfo
  {author} {\bibfnamefont {A.}~\bibnamefont {Buchter}}, \bibinfo {author}
  {\bibfnamefont {F.}~\bibnamefont {Heimbach}}, \bibinfo {author}
  {\bibfnamefont {A.}~\bibnamefont {Fontcuberta~i Morral}}, \bibinfo {author}
  {\bibfnamefont {D.}~\bibnamefont {Grundler}}, \ and\ \bibinfo {author}
  {\bibfnamefont {M.}~\bibnamefont {Poggio}},\ }\href {\doibase
  10.1103/PhysRevB.93.064409} {\bibfield  {journal} {\bibinfo  {journal} {Phys.
  Rev. B}\ }\textbf {\bibinfo {volume} {93}},\ \bibinfo {pages} {064409}
  (\bibinfo {year} {2016})}\BibitemShut {NoStop}%
\bibitem [{\citenamefont {Buchter}\ \emph {et~al.}(2015)\citenamefont
  {Buchter}, \citenamefont {Wölbing}, \citenamefont {Wyss}, \citenamefont
  {Kieler}, \citenamefont {Weimann}, \citenamefont {Kohlmann}, \citenamefont
  {Zorin}, \citenamefont {R\"uffer}, \citenamefont {Matteini}, \citenamefont
  {T\"ut\"unc\"uoglu}, \citenamefont {Heimbach}, \citenamefont {Kleibert},
  \citenamefont {Fontcuberta~i Morral}, \citenamefont {Grundler}, \citenamefont
  {Kleiner}, \citenamefont {Koelle},\ and\ \citenamefont
  {Poggio}}]{buchter_magnetization_2015}%
  \BibitemOpen
  \bibfield  {author} {\bibinfo {author} {\bibfnamefont {A.}~\bibnamefont
  {Buchter}}, \bibinfo {author} {\bibfnamefont {R.}~\bibnamefont {W\"olbing}},
  \bibinfo {author} {\bibfnamefont {M.}~\bibnamefont {Wyss}}, \bibinfo {author}
  {\bibfnamefont {O.~F.}\ \bibnamefont {Kieler}}, \bibinfo {author}
  {\bibfnamefont {T.}~\bibnamefont {Weimann}}, \bibinfo {author} {\bibfnamefont
  {J.}~\bibnamefont {Kohlmann}}, \bibinfo {author} {\bibfnamefont {A.~B.}\
  \bibnamefont {Zorin}}, \bibinfo {author} {\bibfnamefont {D.}~\bibnamefont
  {R\"uffer}}, \bibinfo {author} {\bibfnamefont {F.}~\bibnamefont {Matteini}},
  \bibinfo {author} {\bibfnamefont {G.}~\bibnamefont {T\"ut\"unc\"uoglu}},
  \bibinfo {author} {\bibfnamefont {F.}~\bibnamefont {Heimbach}}, \bibinfo
  {author} {\bibfnamefont {A.}~\bibnamefont {Kleibert}}, \bibinfo {author}
  {\bibfnamefont {A.}~\bibnamefont {Fontcuberta~i Morral}}, \bibinfo {author}
  {\bibfnamefont {D.}~\bibnamefont {Grundler}}, \bibinfo {author}
  {\bibfnamefont {R.}~\bibnamefont {Kleiner}}, \bibinfo {author} {\bibfnamefont
  {D.}~\bibnamefont {Koelle}}, \ and\ \bibinfo {author} {\bibfnamefont
  {M.}~\bibnamefont {Poggio}},\ }\href {\doibase 10.1103/PhysRevB.92.214432}
  {\bibfield  {journal} {\bibinfo  {journal} {Phys. Rev. B}\ }\textbf {\bibinfo
  {volume} {92}},\ \bibinfo {pages} {214432} (\bibinfo {year}
  {2015})}\BibitemShut {NoStop}%
\bibitem [{\citenamefont {Guyader}\ \emph {et~al.}(2012)\citenamefont
    {Guyader}, \citenamefont {Kleibert}, \citenamefont {Rodríguez},
    \citenamefont {Moussaoui}, \citenamefont {Balan}, \citenamefont
    {Buzzi}, \citenamefont {Raabe},\ and\ \citenamefont
    {Nolting}}]{guyader_studying_2012}%
  \BibitemOpen \bibfield {author} {\bibinfo {author} {\bibfnamefont
      {L.~L.}\ \bibnamefont {Guyader}}, \bibinfo {author}
    {\bibfnamefont {A.}~\bibnamefont {Kleibert}}, \bibinfo {author}
    {\bibfnamefont {A.~F.}\ \bibnamefont {Rodr\'iguez}}, \bibinfo
    {author} {\bibfnamefont {S.~E.}\ \bibnamefont {Moussaoui}},
    \bibinfo {author} {\bibfnamefont {A.}~\bibnamefont {Balan}},
    \bibinfo {author} {\bibfnamefont {M.}~\bibnamefont {Buzzi}},
    \bibinfo {author} {\bibfnamefont {J.}~\bibnamefont {Raabe}}, \
    and\ \bibinfo {author} {\bibfnamefont {F.}~\bibnamefont
      {Nolting}},\ }\href {\doibase 10.1016/j.elspec.2012.03.001}
  {\bibfield {journal} {\bibinfo {journal}
      {J. Electron. Spectrosc. Relat. Phenom.}\ }\textbf {\bibinfo
      {volume} {185}},\ \bibinfo {pages} {371} (\bibinfo {year}
    {2012})}\BibitemShut {NoStop}%
\bibitem [{\citenamefont {Vansteenkiste}\ \emph {et~al.}(2014)\citenamefont
  {Vansteenkiste}, \citenamefont {Leliaert}, \citenamefont {Dvornik},
  \citenamefont {Helsen}, \citenamefont {Garcia-Sanchez},\ and\ \citenamefont
  {Van~Waeyenberge}}]{vansteenkiste_design_2014}%
  \BibitemOpen
  \bibfield  {author} {\bibinfo {author} {\bibfnamefont {A.}~\bibnamefont
  {Vansteenkiste}}, \bibinfo {author} {\bibfnamefont {J.}~\bibnamefont
  {Leliaert}}, \bibinfo {author} {\bibfnamefont {M.}~\bibnamefont {Dvornik}},
  \bibinfo {author} {\bibfnamefont {M.}~\bibnamefont {Helsen}}, \bibinfo
  {author} {\bibfnamefont {F.}~\bibnamefont {Garcia-Sanchez}}, \ and\ \bibinfo
  {author} {\bibfnamefont {B.}~\bibnamefont {Van~Waeyenberge}},\ }\href
  {\doibase 10.1063/1.4899186} {\bibfield  {journal} {\bibinfo  {journal} {AIP
  Adv.}\ }\textbf {\bibinfo {volume} {4}},\ \bibinfo {pages} {107133}
  (\bibinfo {year} {2014})}\BibitemShut {NoStop}%
\end{thebibliography}
\end{document}